\documentclass[%
 reprint,
 amsmath,amssymb,
 aps,
]{revtex4-2}

\usepackage{amsmath}
\usepackage{amssymb}
\usepackage{commath}
\usepackage{mathtools} 
\usepackage{times}
\usepackage{siunitx}
\usepackage{hyperref}
\hypersetup{colorlinks=true, linkcolor=[RGB]{0,0,255}, citecolor=[RGB]{0,0,255}, urlcolor=blue}

\usepackage{doi}
\usepackage{orcidlink}
\usepackage[utf8]{inputenc}
\usepackage{graphicx}
\usepackage{xcolor}
\usepackage[switch]{lineno}

\usepackage{graphicx}
\usepackage{dcolumn}
\usepackage{bm}

\begin{document}

\preprint{APS/123-QED}

\title{Two-phase Temperature Reconstruction in Ice-Water Systems}

\author{Zhukun Wang}
\affiliation{Department of Earth and Environmental Science, University of Pennsylvania, Philadelphia, PA19104, USA.}

\author{Daisuke Noto}%
\affiliation{%
 Department of Earth and Environmental Science, University of Pennsylvania, Philadelphia, PA19104, USA.\\Faculty of Engineering, Hokkaido University, Sapporo, 060-8628, Japan.\\
 Japan Agency for Marine-Earth Science and Technology (JAMSTEC), Yokosuka, 236-0001, Japan.
}%

\author{Douglas J. Jerolmack}
\affiliation{Department of Earth and Environmental Science, University of Pennsylvania, Philadelphia, PA19104, USA.\\
Department of Mechanical Engineering and Applied Mechanics, University of Pennsylvania, Philadelphia, Pennsylvania 19104, USA.}

\author{Hugo N. Ulloa}\thanks{Corresponding Author: Hugo N. Ulloa. Email: ulloa@sas.upenn.edu}%
\affiliation{Department of Earth and Environmental Science, University of Pennsylvania, Philadelphia, PA19104, USA.\\
}

\date{\today}

\begin{abstract}
What controls heat transport when liquid water interacts with ice? 
Answering this question is essential for understanding water bodies undergoing phase change. 
Yet experimental progress remains limited by the lack of a minimally invasive methodology for simultaneously resolving temperature fields in coupled liquid water and non-isothermal ice systems. 
Here, we introduce a physics-based data-assimilation method for reconstructing temperature fields in buoyancy-driven flows interacting with non-isothermal ice.
Particle tracking velocimetry provides the liquid velocity field, while thermal and kinematic boundary conditions constrain the inverse problem.
The method couples advection--diffusion in liquid water with conduction in ice to reconstruct simultaneous mean temperature fields and quantify heat transport across the water--ice interface, while remaining minimally invasive and compatible with free-surface systems.
We demonstrate the method in laboratory experiments in which the temperature range across the liquid water in contact with ice drives cabbeling-induced convection.
This framework enables investigation of coupled thermo-fluid dynamics in cryospheric aquatic systems, including heat exchange at the ice-water interface and liquid-phase energetics, with broader applications to phase-change processes in food and energy industries.

\end{abstract}

\maketitle

\section{Introduction}\label{sec:introduction}

Solid--liquid water systems are ubiquitous in both natural environments and engineered technologies, spanning a wide range of length scales, geometries, and forcing conditions. In nature, they include floating ice bodies such as icebergs and ice floes, supraglacial meltwater bodies, subglacial and proglacial lakes, and seasonally ice-covered lakes \citep[e.g.,][]{wells2008circulation,kirillin2015axisymmetric,sugiyama2019underwater,fernandez2021laboratory,nash2024turbulent,ryan2025meltwater,rabaux2026hysteretic}. Their engineered counterparts include ice-based thermal-energy storage, ice-source heat pumps, supercooled-water ice generation, and pumpable ice-slurry systems used in food processing \citep[e.g.,][]{selvnes2021review,ahn2022performance,li2022experimental,fang2022exploring,tang2023optimization,yedmel2024experimental,cheng2025dynamic}. Despite this apparent diversity, these systems are governed by the same tightly coupled transport problem: heat must be redistributed through media with fundamentally different transport mechanisms---conduction in the solid and advection--diffusion in the liquid---while being exchanged across a moving phase boundary, controlling melting or freezing, and the thermo-fluid dynamics of the liquid phase.

Ice--liquid water (hereafter ice--water) systems therefore provide canonical settings in which heat transport, phase change and buoyancy-driven flow are strongly coupled \citep{bushuk2019ice,wang2021equilibrium,wang2021growth,hester2021aspect,weady2022anomalous,mccutchan2023experimental,johnson2025shape,bellincioni2025melting,noto2026melting,perry2026pof}. Temperature gradients drive heat conduction through the ice and diffusion and buoyancy-driven advection through the water, whereas the imbalance between the heat fluxes arriving at the interface governs its local advance or retreat. These processes motivate three fundamental questions: (i) How is heat distributed across the solid and liquid phases? (ii) How does ice geometry organize the surrounding circulation and heat transport? (iii) And how does that circulation regulate interfacial heat exchange and phase change? Although these questions provide the physical motivation for this study, the present work addresses the experimental capability required to investigate them. 

Answering these questions requires simultaneous estimates of temperature in ice and water, liquid velocity, and interfacial heat flux without substantially disturbing the flow, interface geometry, or thermal boundary conditions. Existing optical thermometry methods provide detailed temperature measurements in liquids, but their application to ice--water systems remains limited by large temperature differences between the phases, free surfaces, irregular interfaces, and phase change. The principal approaches are laser-induced fluorescence (LIF) and liquid-crystal thermometry (LCT) \citep{someya2021particle}. LIF infers temperature from the calibrated fluorescence intensity of temperature-sensitive dyes \citep{sakakibara2004measurement,voulgaropoulos2022simultaneous,kashanj2023application}. Its accuracy can deteriorate through photobleaching, which changes the intensity--temperature calibration over time \citep{eghtesad2024state}, and through spatial variations in dye concentration. LCT instead uses thermochromic liquid-crystal particles as tracers for simultaneous velocity and temperature measurements \citep{dabiri2009digital,abdullah2010basics}, with particle color mapped to temperature over a calibrated interval \citep{kaufer2023volumetric}. Recent neural-network and physics-informed extensions have enabled three-dimensional thermofluid reconstruction and applications to solidifying flows \citep{anders2020simultaneous,toscano2025aivt}. However, the useful temperature interval of thermochromic liquid crystals is often limited to only a few kelvin, restricting their application to systems spanning subfreezing ice temperatures and substantially warmer liquid water.

More fundamentally, LIF and LCT measure temperature only in regions containing the dye or tracer particles. During solidification, the advancing ice interface typically excludes or redistributes dissolved substances and suspended particles \citep[e.g.,][]{middleton2022relative,olsthoorn2022salt}. The resulting experimental domain therefore contains a seeded liquid phase and a solid phase that is largely inaccessible to tracer-based optical thermometry. Direct optical measurements may consequently constrain temperature in the water but cannot generally provide the corresponding temperature field within the ice. They therefore cannot, on their own, quantify heat transport through both phases or evaluate the heat-flux balance across the ice--water interface.

An alternative strategy is to infer temperature from measured fluid motion and known thermal boundary conditions rather than measuring the scalar field directly. \citet{bauer2022assimilation}, for example, combined Particle Image Velocimetry (PIV) with direct numerical simulation through data assimilation to recover instantaneous temperature fields \citep[see also][]{volk2025pinn,barta2025temperature}. Such approaches can recover detailed thermofluid information but require iterative coupling between measurements and a numerical model, together with careful selection of model and optimization parameters. A more direct physics-based approach was introduced by \citet{noto2023reconstructing}, who reconstructed temperature in Boussinesq flows from PIV measurements by solving the heat advection--diffusion equation subject to prescribed thermal boundary conditions. Related implementations have produced physically consistent temperature reconstructions in stratified, plume-driven, and convective flows \citep{noto2023stratified,noto2024plume,yanagisawa2024quasi,ulloa2025convection,noto2025convective}. These methods demonstrate that velocity measurements can constrain otherwise inaccessible thermal fields, but they have primarily been formulated for single-phase liquid domains. They do not simultaneously reconstruct conduction in a non-isothermal solid, advection--diffusion in the adjoining liquid, and heat-flux continuity or imbalance across an experimentally measured solid--liquid interface.

\begin{figure*}
\centering\includegraphics[width=1\linewidth]{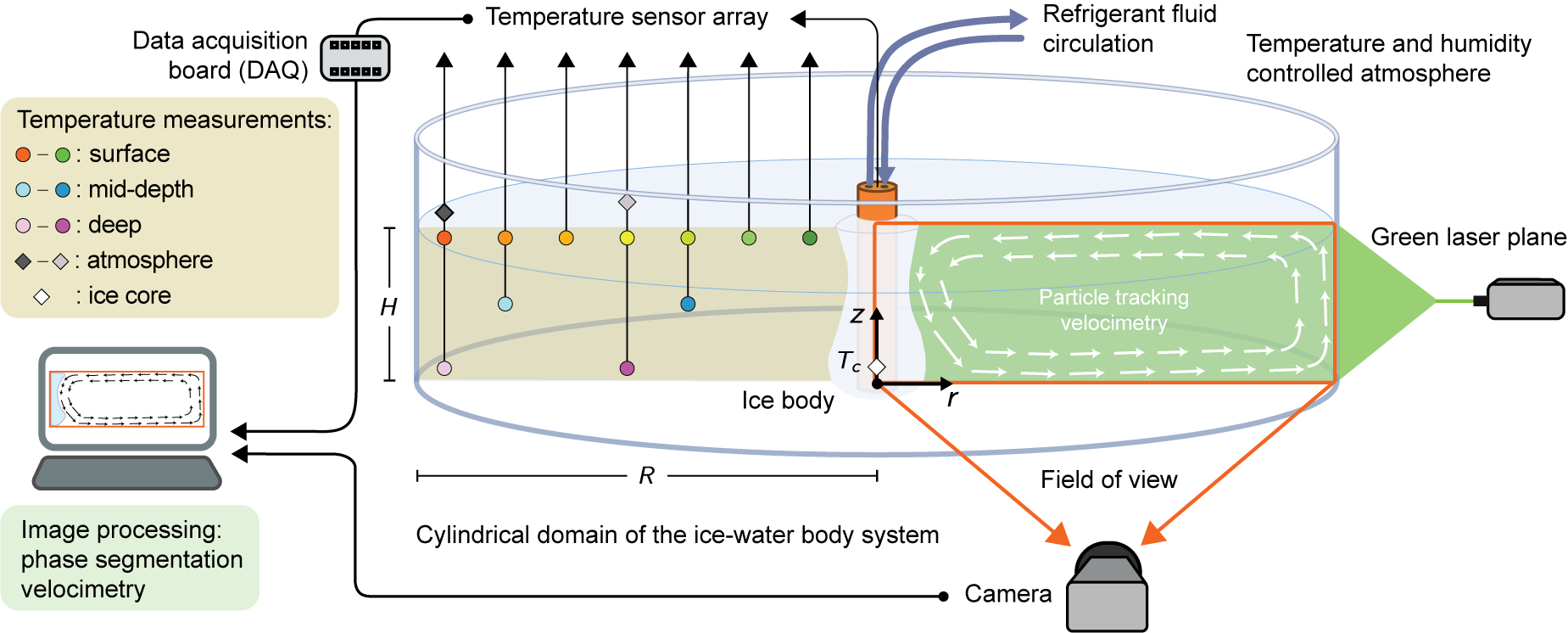}
  \caption{Schematic of the experimental setup. A cylindrical water domain is enclosed within a square tank (not shown) to avoid undesired optical refraction. At the center of the domain, an ice body is formed around a copper tube through which a temperature-controlled refrigerant is circulated. A laser sheet is introduced along a radial plane to quantify the convective flow using PTV. The shaded region on the left indicates the $r$-$z$ plane in which eleven small-scale temperature sensors are deployed within the water, with their locations indicated by colored circles. The colors provide a direct correspondence between sensor locations and the measurements reported in Fig.~\ref{fig:8}, Fig.~\ref{fig:Appendix_fig1}, and Fig.~\ref{fig:Appendix_fig2}. Additionally, two sensors record the ambient air temperature, while one sensor monitors the temperature inside the copper tube.}
  \label{fig:1}
\end{figure*}

Here, we develop and validate a physics-based data-assimilation framework for two-phase temperature reconstruction in an ice--water system with an air--water free surface. The method combines Particle Tracking Velocimetry measurements in the liquid, an experimentally detected ice--water interface, sparse temperature measurements, and prescribed or parameterized thermal boundary conditions. Temperature in the liquid is reconstructed by solving the steady advection--diffusion equation using the measured velocity field, while temperature in the ice is obtained independently from the steady conduction equation. The two phase-specific solutions are coupled through the measured interface geometry, the prescribed freezing temperature, and an interfacial heat-flux balance used to determine the unknown surface-temperature parameters. Independent thermistor measurements are then used to assess the reconstructed fields. The resulting framework provides simultaneous, minimally invasive estimates of temperature and heat transport in both phases for irregular, non-isothermal ice geometries under quasi-steady experimental conditions. It thereby supplies the thermal information needed to investigate how ice geometry, buoyancy-driven circulation, and interfacial heat transfer interact in laboratory ice--water systems.

The article is organized as follows. Section~\ref{sec:methods} introduces the principle of the method and describes the experimental setup, governing equations, and reconstruction procedure. Section~\ref{sec:results} presents reconstructed fields for cabbeling-induced convection in water bodies interacting with non-isothermal ice of irregular geometry and validates the method against independent temperature measurements. Section~\ref{sec:discussion} examines the assumptions, applicability, and limitations of the framework. Finally, Sect.~\ref{sec:summary} summarizes the main methodological advances and outlines potential applications.


\section{Methods}\label{sec:methods}

\subsection{Principle}\label{sec:principle}

We consider a laboratory-scale ice-water system subject to controlled thermal forcing (Fig.~\ref{fig:1}). The liquid is open to a dark atmosphere, which supplies heat at the air--water interface, while the immersed ice extracts heat through the ice--water interface. These two exchanges set the temperature field in both phases and drive the coupled thermo-fluid response. 

In the liquid phase, temperature evolves according to heat transport equation
\begin{equation}\label{eq:heat_liquid}
\partial_{t}T + {\bm u}\cdot \nabla T = \kappa_{\rm w}\,\nabla^{2}T,
\end{equation} 
where $T$ represents the fluid temperature,  ${\bm u}$ denotes the velocity field, and $\kappa_{\rm w}$ the thermal diffusivity of water. In the ice, heat transport is conductive,
\begin{equation}\label{eq:heat_solid}
\partial_{t}T = \kappa_{\rm i}\,\nabla^{2}T,
\end{equation}
where $\kappa_{\rm i}$ is the thermal diffusivity of ice.

Following \citet{noto2023reconstructing}, we reconstruct the mean temperature field $\overline{T}$ from the mean velocity field $\overline{\bm u}$ at quasi-steady state (QSS). 
This state is expected to hold when the temperature and velocity fields remain approximately stationary, and the ice--water interface changes negligibly over the averaging time interval.

At QSS, we decompose the fields as
$$
{\bm u}({\bm x},t)=\overline{\bm u}({\bm x})+{\bm u}'({\bm x},t),\qquad
T({\bm x},t)=\overline{T}({\bm x})+T'({\bm x},t),
$$
and we assume that fluctuations ${\bm u}'$ and $T'({\bm x},t)$ have zero mean, and that turbulent or unsteady heat fluxes are small compared with the mean advective and diffusive fluxes,
$$
||\overline{\bm u}\,\overline{T}|| \gg ||\overline{{\bm u}'T'}||,\qquad
||\kappa\nabla\overline{T}|| \gg ||\overline{{\bm u}'T'}||.
$$
The governing equations therefore reduce to steady advection--diffusion in the liquid,
\begin{equation}\label{eq:heat_red_liquid}
\overline{\bm u}\cdot \nabla \overline{T} =  \kappa_{\rm w}\,\nabla^{2}\overline{T},
\end{equation}
and Laplace conduction in the ice,
\begin{equation}\label{eq:heat_red_solid}
\nabla^{2}\overline{T}=0.
\end{equation}

Because the liquid and solid subdomains are governed by different heat-transport mechanisms and have distinct material properties, we reconstruct their temperature fields separately and couple them through the experimentally identified ice--water interface.

First, the mean temperature field in the solid phase is reconstructed by discretizing Eq.~\ref{eq:heat_red_solid} and imposing the corresponding boundary conditions. The ice core is maintained by a chiller at a prescribed temperature, $T_{\rm c}<T_{\rm i}=0^\circ{\rm C}$, while the ice--water interface is held at the freezing temperature, $T_{\rm i}=0^\circ{\rm C}$. All remaining boundaries of the solid domain are treated as adiabatic, such that $\partial \overline{T}/\partial n=0$.

Second, for reconstructing the mean temperature in the liquid phase, we obtain the mean velocity field $\overline{\bm u}$ from Particle Tracking Velocimetry (PTV). The temperature field is then reconstructed by discretizing Eq.~\ref{eq:heat_red_liquid}, assimilating $\overline{\bm u}$, and imposing velocity and temperature boundary conditions. 

The kinematic boundary conditions are no slip on solid boundaries, including the ice-water interface, and free slip at the air--water interface. Consistently, the ice--water interface is held at the freezing temperature, $T_{\rm i}=0^\circ{\rm C}$, whereas the remaining no-ice solid boundaries are treated as adiabatic, such that $\partial \overline{T}/\partial n=0$. In contrast, the temperature at the air-water interface is not prescribed and, \textit{a~priori}, unknown. Unlike previous single-phase reconstructions \citep{noto2023reconstructing,noto2024plume,ulloa2025convection,noto2025convective}, this open boundary is set by the thermal coupling between the atmosphere, liquid, and ice. We treat the atmosphere as a constant-temperature reservoir at $T_{\rm a}$. Hence, at QSS, the liquid heat budget requires the heat entering through the air--water interface to balance the heat extracted at the ice--water interface,
\begin{equation}\label{eq:heat_budget}
\Bigg|\oint_{\partial\Omega_{\rm iw}}k_{\rm w}\nabla T_{\ell}\cdot\hat{\bm{n}}_{\ell}\,{\rm d}S\Bigg|
=
\Bigg|\oint_{\partial\Omega_{\rm aw}}k_{\rm w}\nabla T_{\ell}\cdot\hat{\bm{n}}_{\rm aw}\,{\rm d}S\Bigg|.
\end{equation}
For a stationary ice--water interface, the local conductive fluxes must also balance across the interface,
\begin{equation}\label{eq:heat_flux_ice_water}
 k_{\rm i}\nabla T_{\rm s}\cdot \hat{\bm n}_{\rm s}
+
 k_{\rm w}\nabla T_{\ell}\cdot \hat{\bm n}_{\ell}=0,
\end{equation}
where $\hat{\bm n}_{\rm s}=-\hat{\bm n}_{\ell}$ are the outward normals of the solid and liquid phases. The unknown air--water temperature distribution, $T_{\rm aw}(\partial\Omega_{\rm aw})$, is therefore inferred by enforcing these heat-budget constraints.

The surface temperature function must satisfy $T_{\rm aw}=T_{\rm i}$ at the ice--water--air contact line and $\partial T_{\rm aw}/\partial n=0$ at the outer boundary. We write 
$$T_{\rm aw}=\mathcal{F}(\Delta T,\delta),$$
where $\Delta T=T_{\rm R}-T_{\rm i}$, and $T_{\rm R}$ is the surface temperature at the outer boundary. The parameter $\delta$ is the characteristic surface thermal-boundary-layer length --- over which $\Delta T$ is established. The functional form of $\mathcal{F}$ and the unknown parameters $\Delta T$ and $\delta$ are chosen to minimize the QSS heat-budget residual. 

\begin{figure*}[ht]
\centering\includegraphics[width=1\linewidth]{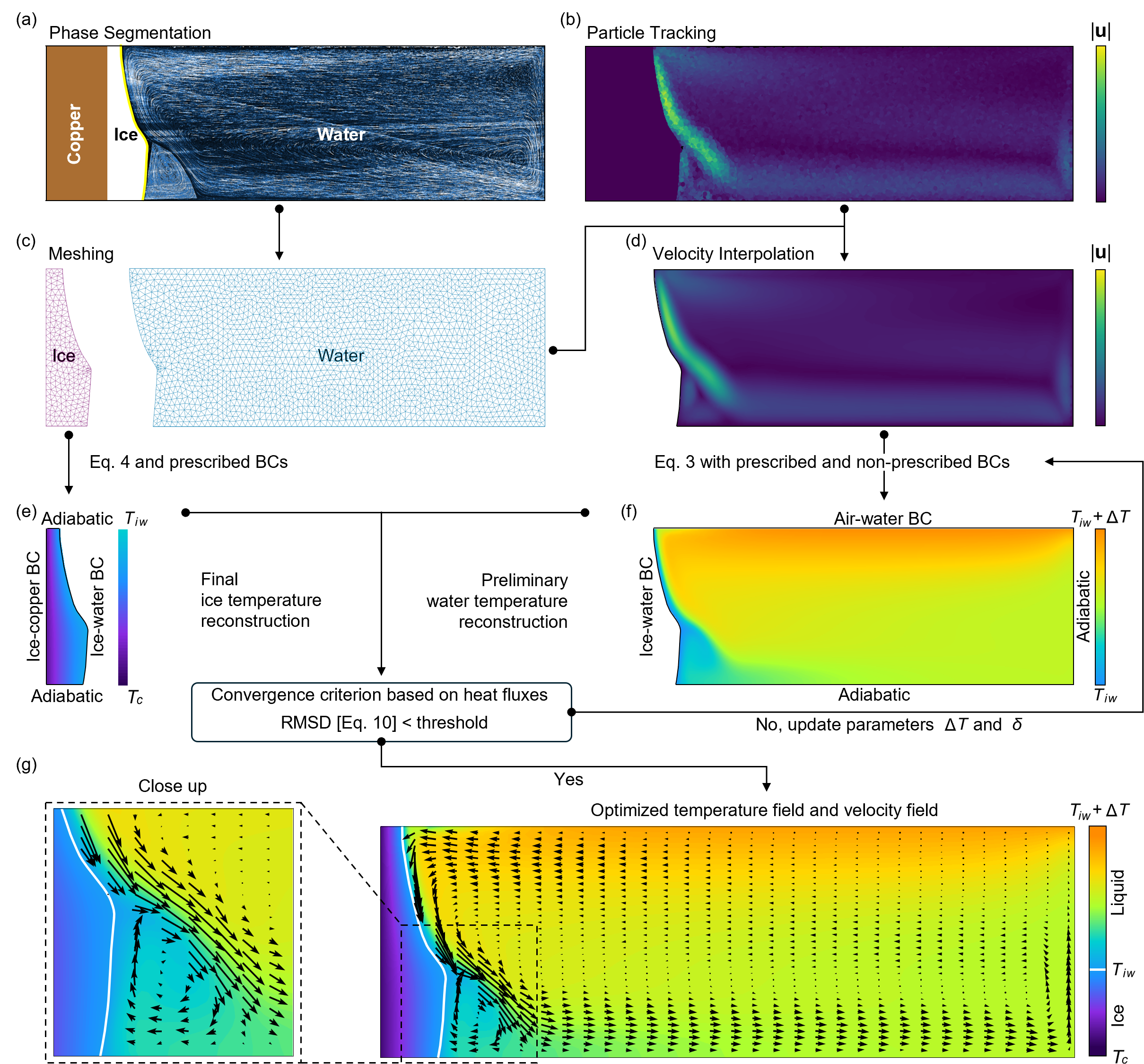}
  \caption{Overview of the two-phase temperature reconstruction: (a) phase-segmentation and pathline imaging under QSS; (b) velocity data obtained from PTV; (c) meshing of the liquid and solid phases; (d) interpolation of the PTV-derived velocity field onto the computational mesh; 
  (e) final ice temperature reconstruction through solving the Laplace equation \ref{eq:heat_red_solid}; (f) assimilation of velocity and specification of boundary conditions to integrate the advection-diffusion equation \ref{eq:heat_red_liquid} in the liquid phase to reconstruct the temperature field; (g) output: thermally driven flow field and temperature field in the liquid and solid phases. 
  Between (f) and (g), there's the parameter seeking procedure where the optimum parameters are searched for so that the heat budget is closed with the minimum RMSD. }
  \label{fig:2}
\end{figure*}

\subsection{Experimental setup}\label{sec:experiment}

Ice--water interfaces can evolve into complex shapes. To test the temperature reconstruction, principle under a controlled geometry, we designed an annular ice--water system in which the interface evolves \textit{quasi} axisymmetrically (Fig.~\ref{fig:1}). This configuration reduces three-dimensional (3D) effects and provides a well-defined setting for linking fluid velocity, interface geometry, and thermal boundary conditions.

The setup consists of an open-ended acrylic cylinder of internal radius $R=171~{\rm mm}$ placed inside an open-top square acrylic tank of height $150~{\rm mm}$. A copper cylinder of outer radius $R_{\rm c}=20.5~{\rm mm}$ is mounted concentrically within the acrylic cylinder, forming an annular water domain bounded radially by the copper surface and the outer acrylic wall (Fig.~\ref{fig:1}). The transparent acrylic provides optical access for side-view imaging.

The copper cylinder contains an internal cavity connected to a thermostatic bath, which circulates temperature-controlled refrigerant and maintains the copper surface at a prescribed temperature $T_{\rm c}<T_{\rm i}$. The acrylic tank is insulated with 50~mm-thick extruded polystyrene (XPS) foam to minimize heat exchange through the outer walls. The full setup is enclosed in a 50~mm-thick XPS housing mounted on the optical table, suppressing diurnal temperature variations and providing a quasi-steady atmospheric environment. The air temperature inside the housing is maintained at $T_{\rm a}\approx 20^{\circ}$C by a PID-controlled Peltier module. These controls approximate a constant-temperature copper boundary, adiabatic acrylic walls, and an air--water temperature distribution to be inferred from the heat-budget constraint described above (\ref{sec:principle}).

Each experiment begins by filling the annular domain with deionized water initially at $T_{\rm a}$. Ice then grows radially outward from the cooled copper surface, establishing a primarily radial temperature gradient and a buoyancy-driven flow with no imposed azimuthal preference. The space between the annular cylinder and the surrounding square tank is also filled with water to reduce refractive-index mismatch and optical distortion during side-view imaging.

Velocity and ice-water interface position are obtained in a vertical $r$--$z$ plane. The liquid is seeded with tracer particles of mean diameter $20~{\rm \mu m}$ and illuminated periodically by a 1~mm-thick green laser sheet (Fig.~\ref{fig:1}). A CMOS camera records particles motion for PTV and the ice--water interface for solid-liquid phase segmentation. Image acquisition parameters are selected according to flow characteristics. The spatial resolution is chosen to adequately resolve the flow structures of interest, while the frame rate is selected such that particle displacements between consecutive frames remain within the measurable range, typically from subpixel resolution to approximately 10 pixels.

To preserve insulation while enabling imaging, an automated stepper motor briefly removes the XPS panel facing the camera during each acquisition and replaces it immediately afterward. Images are collected during this interval, which typically lasts 30~s.

Temperature is measured independently for validation and boundary-condition monitoring. An array of thermistors is installed at fixed underwater locations on the side opposite the imaging plane (Fig.~\ref{fig:1}). The thermistors are color-coded to distinguish their individual measurement locations, with the same color coding used to compare measured and reconstructed temperatures at each location. Under axisymmetric conditions, these measurements provide independent error estimates for the reconstructed temperature fields. Three thermocouples provide reference temperatures: one at the center of the copper cylinder to monitor $T_{\rm c}$ and two in the air near the air--water interface to monitor $T_{\rm a}$. 

Image acquisition, laser operation, motor motion, and temperature sensing are synchronized and automated, allowing measurements without manual interaction with the enclosed setup.

Control experiments without ice were conducted to determine whether laser-induced heating produces measurable background convection and to isolate the role of the ice in shaping the observed flow. These experiments showed that the laser-induced circulation is weak and negligible relative to the buoyancy-driven flow observed in the presence of ice (see Appendix~\ref{sec:appendix_A}).

\subsection{Two-phase temperature reconstruction in an axisymmetric domain}\label{sec:theory}

Once the liquid temperature measurements and ice volume reach QSS, the reconstruction starts from image acquisition. Phase segmentation identifies the ice--water interface in the $r$--$z$ plane (Fig.~\ref{fig:2}a), and an in-house PTV code \citep{noto2021developing} resolves the liquid velocity field, including the vicinity of the irregular ice boundary (Fig.~\ref{fig:2}b).

Since the flow is \textit{quasi} axisymmetric, the reduced heat equations \ref{eq:heat_liquid} and \ref{eq:heat_solid} are written in cylindrical coordinates as
\begin{equation}\label{eq:heat_cyl_liquid}
\kappa_{\rm w}\,\left(\frac{\partial^2 \overline{T}}{\partial r^2} 
+ \frac{1}{r} \frac{\partial \overline{T}}{\partial r}  
+ \frac{\partial^2 \overline{T}}{\partial z^2}\right)
= \overline{u_r}\,\frac{\partial \overline{T}}{\partial r} + \overline{u_z}\,\frac{\partial \overline{T}}{\partial z},
\end{equation}
for the liquid and
\begin{equation}\label{eq:heat_cyl_solid}
\kappa_{\rm i}\,\left(\frac{\partial^2 \overline{T}}{\partial r^2} 
+ \frac{1}{r} \frac{\partial \overline{T}}{\partial r}  
+ \frac{\partial^2 \overline{T}}{\partial z^2}\right)
= 0,
\end{equation}
for the ice. For the numerical integration, we use reference values for thermal diffusivities representative of the temperature ranges observed in the experiments. For liquid water, we take $\kappa_{\rm w}=1.34\times 10^{-7}~\rm m^{2}\,s^{-1}$, and for ice, we take $\kappa_{\rm i}=1.02\times 10^{-6}~\rm m^{2}\,s^{-1}$.

The air--water temperature, $T_{\rm aw}(r)$, is not known \textit{a priori}. It is constrained by the contact-line condition $T_{\rm aw}(r_{\rm iw})=T_{\rm i}$ and the outer no-flux condition $\left.\partial T/\partial r\right|_{r=R}=0$. We model the radial temperature distribution of this boundary as
\begin{equation}\label{eq:Taw_radial}
T_{\rm aw}(r) = T_{\rm i} + \Delta T \tanh \left(\frac{r-r_{\rm iw}}{\delta}\right),
\end{equation}
where $\Delta T=T_{\rm R}-T_{\rm i}$ and $\delta$ are constrained by the experimental conditions, near-surface thermistor measurements (Appendix~\ref{sec:appendix_B}; Fig.~\ref{fig:Appendix_fig2}) and optimized as described in Sect.~\ref{sec:param_opt}. A hyperbolic tangent function is adopted because it naturally satisfies the required thermal boundary conditions. At the ice--water interface, located at $r=r_{\rm iw}$, the temperature is fixed at $T=T_{\rm i}=0^\circ\mathrm{C}$. At the outer cylindrical wall, $r=R$, the temperature approaches $T_{\rm R}$ while satisfying an approximately adiabatic condition, $T(R)\approx T_R$ and $\left. \partial T/\partial r\right|_{r=R}\approx 0.$

The solid- and liquid-phase temperature fields are computed numerically in the $r$--$z$ plane using FreeFEM~\citep{hecht2012new}, an open-source finite-element solver for partial differential equations. The experimentally detected ice--water interface is represented as a piecewise-linear contour constructed from an ordered set of discrete interface points. This contour is connected to the corresponding physical boundaries of the experimental domain to define closed computational subdomains for the ice and liquid water. Each subdomain is then discretized using an unstructured mesh of triangular finite elements, with the experimentally determined interface forming their shared boundary (Fig.~\ref{fig:2}c).

In the ice, Eq.~\ref{eq:heat_cyl_solid} is solved with fully prescribed boundary conditions. The ice--water interface is fixed at $T_{\rm iw}=0^\circ{\rm C}$, and the ice--copper interface is fixed at the measured temperature $T_{\rm c}<T_{\rm i}$. Ice boundaries in contact with the tank bottom or XPS are treated as adiabatic. When the ice reaches the symmetry axis (i.e. $r=0$ and anywhere between the bottom $z=0$ and the air-water interface at $z=H$), a no-flux condition is imposed there to respect symmetry. These conditions uniquely determine the conductive temperature field in the ice (Fig.~\ref{fig:2}e).

In the liquid, the PTV-resolved mean velocity field is interpolated onto the mesh (Fig.~\ref{fig:2}b,d) and assimilated into Eq.~\ref{eq:heat_cyl_liquid}. No-slip conditions are imposed on solid boundaries, including the ice--water interface, and a shear-free condition is imposed at the air--water interface. When the liquid domain includes the symmetry axis, shear-free and no-flux conditions are applied at $r=0$. The temperature is prescribed at the ice--water interface, $T=T_{\rm i}$, and at the air--water interface using Eq.~\ref{eq:Taw_radial}. Acrylic walls, including bottom and side boundaries, are treated as adiabatic; the same condition is used in floating-ice configurations where water separates the ice from the tank bottom.

The weak forms of Eqs.~\ref{eq:heat_cyl_liquid} and \ref{eq:heat_cyl_solid}—their integral formulations obtained by weighting the governing equations with test functions—are solved in FreeFEM to reconstruct the ice temperature field and a provisional liquid temperature field for each candidate parameter pair $(\Delta T,\delta)$. The liquid solution is accepted only when the corresponding surface-temperature parameters satisfy the ice--water heat-flux balance in Eq.~\ref{eq:heat_flux_ice_water} (Fig.~\ref{fig:2}d,f). The optimization procedure used to identify these parameters is described next.

\subsection{Parameter optimization}\label{sec:param_opt}

The surface-temperature model in Eq.~\ref{eq:Taw_radial} contains two unknown parameters, $\Delta T$ and $\delta$. These parameters must be optimized to close the air--water thermal boundary condition and reconstruct the liquid temperature field.

The optimization proceeds in two stages. First, a coarse grid search is performed over the admissible parameter space. The temperature amplitude $\Delta T$ is varied between the freezing temperature, $T_{\rm i}$, and the ambient air temperature, $T_{\rm a}$. A first approximation for $\delta$ is estimated from the seven near surface thermistor array (Fig.~\ref{fig:1}, Fig~\ref{fig:Appendix_fig2}). To ensure a conservative search, the upper bound for $\delta$ is set to three times this length. The grid spacings for $\Delta T$ and $\delta$ are chosen to match, respectively, the thermistor temperature resolution and the computational mesh spacing.

For each candidate pair $(\Delta T,\delta)$, Eq.~\ref{eq:Taw_radial} is imposed at the air--water interface and the liquid heat equation, Eq.~\ref{eq:heat_cyl_liquid}, is solved to obtain a tentative temperature field. This field is then combined with the precomputed conductive temperature field in the ice. For a stationary ice--water interface, the two-phase solution must satisfy both the global liquid heat budget, Eq.~\ref{eq:heat_budget}, and the local interfacial flux balance in Eq.~\ref{eq:heat_flux_ice_water}. The heat flux is evaluated segment-wise on both sides of the segmented ice--water interface.

In practice, residual flux imbalances arise from numerical truncation, measurement uncertainty, and departures from exact steady state. We quantify them using the root-mean-square deviation of the local heat-flux mismatch,
\begin{equation}\label{eq:RMSD_heat_budget}
{\rm RMSD} =
\sqrt{
\frac{1}{\mathcal{L}}
\oint_{\partial\Omega_{\rm iw}}
\left(
k_{\rm w}\frac{\partial T_{\rm w}}{\partial n}\Big|_{\rm w}
+
k_{\rm i}\frac{\partial T_{\rm i}}{\partial n}\Big|_{\rm i}
\right)^2
{\rm d}S
}.
\end{equation}
Here, $\mathcal{L}$ is the total length of the ice--water interface locus, $\partial\Omega_{\rm iw}$, obtained from phase segmentation in the $r$--$z$ plane. The derivatives are projected along the local interface normal, and $k_{\rm w}$ and $k_{\rm i}$ are the thermal conductivities of water and ice. The RMSD is computed for every $(\Delta T,\delta)$ pair, and the minimum defines the coarse optimum.

In the second stage, the search is repeated on a finer grid centered on the coarse optimum. The refined pair is accepted as optimal when the minimum RMSD reaches a $\mathcal{O}(1\%)$ tolerance; otherwise, the local grid is further refined. The resulting values of $\Delta T$ and $\delta$ in Eq.~\ref{eq:Taw_radial} constrain the air--water temperature boundary condition and, consequently,the corresponding liquid temperature field.

\subsection{Quality checks for optimal solution}

\subsubsection{RMSD-based check using ice--water interface displacement}

We first assess the reconstruction by comparing two independent estimates of ice--water interface displacement. The first is inferred from the residual heat-flux imbalance of the reconstructed temperature field; the second is measured directly from segmented images. This test determines whether the minimum RMSD implies an interface displacement comparable to the observed interface variability and detection uncertainty at QSS.

The minimum RMSD is converted into an equivalent residual interface displacement over the measurement window. Because the RMSD represents an integrated heat-rate residual, it is divided by the interfacial area per unit azimuthal length, $\mathcal{L}$, to define an equivalent residual heat flux,
$$q_{\rm res}=\frac{\mathrm{RMSD}}{\mathcal{L}}.$$
Using the Stefan condition \citep[e.g.,][]{pegler2021convective}, this residual heat flux is expressed as an equivalent interface-normal velocity,
$$V_{\rm res}=\frac{q_{\rm res}}{\rho_{\rm ice}\,L_{\rm f}},$$
where $\rho_{\rm ice}$ is the ice density and $L_{\rm f}$ is the specific latent heat of fusion of pure water. Notice that freezing releases the same latent heat in magnitude as the latent heat absorbed during melting; its sign is accounted for through the adopted heat-flux and interface-velocity conventions. The corresponding equivalent interface displacement over the measurement interval $\tau$ is
\begin{equation}\label{eq:l_res}
    \ell_{\rm res}= V_{\rm res}\tau .
\end{equation}

The measured interface displacement is evaluated by comparing two segmented ice--water interfaces within the QSS regime. The earlier interface is first discretized into linear segments. For each segment, a normal vector is constructed at its midpoint and extended until it intersects the later interface. The distance between the midpoint and the corresponding intersection, measured along the normal direction, defines the local interface displacement $\ell_i$. Repeating this procedure for all segments yields a distribution of local displacements, whose mean value is taken as the characteristic measured displacement, $\overline{\ell}_{\tau}$.

We then include interface-detection uncertainty. The minimum segmentation uncertainty is approximately two pixels, corresponding to a possible $\pm 1$ pixel shift when the true interface lies between adjacent pixels. The image-based estimate of interface-position variability is therefore
\begin{equation}\label{eq:l_exp}
    \ell_{\rm exp}=\overline{\ell}_{\tau}+\ell_{\rm det},
\end{equation}
where $\ell_{\rm det}$ is the detection uncertainty. The reconstruction passes this quality check when
$$    \mathcal{O}(\ell_{\rm res}) \lesssim \mathcal{O}(\ell_{\rm exp}).$$

In this case, the residual heat-flux imbalance is within the experimental uncertainty of the observed interface fluctuation, and the optimal pair $(\Delta T,\delta)$ is accepted. If $\ell_{\rm res}$ is significantly larger than $\ell_{\rm exp}$, the reconstructed heat flux is inconsistent with the QSS assumption, indicating that the reconstruction procedure is no longer valid. A robust quasi-steady state must therefore be re-established before the thermal field can be reconstructed. In practice, the most reliable approach is to restart the experiment and repeat the reconstruction procedure from phase segmentation, PTV, meshing, and velocity interpolation onward (Fig.~\ref{fig:2}a--d).

\subsubsection{Temperature check using \textit{in situ} measurements}

We next evaluate the reconstructed liquid temperature field using independent \textit{in situ} measurements (Fig.~\ref{fig:1}). This comparison accounts for the different sampling: the sensors provide time-averaged local measurements, whereas the reconstruction provides a spatially resolved temperature field.

At QSS, the temperature recorded by each sensor `$i$' is averaged over the measurement window to obtain $\overline{T}^{\rm exp}_{i}$. The corresponding reconstructed value, $\langle T^{\rm mod}_{i}\rangle$, is computed by averaging the liquid-phase temperature over a small region around the sensor location. This averaging accounts for finite sensor size and positioning uncertainty; here, the region is circular with a diameter equal to three times the sensor thickness.

Agreement between $\overline{T}^{\rm exp}_{i}$ and $\langle T^{\rm mod}_{i}\rangle$ provides an independent validation of the liquid-phase temperature reconstruction.


\section{Demonstration}\label{sec:results}

We apply the workflow summarized in Fig.~\ref{fig:2} to three representative ice--water configurations: (a) full-depth ice, (b) intermediate-depth ice, and (c) surface-confined ice. These cases systematically vary the degree of ice submergence, providing a stringent test of the method across distinct geometric configurations. A representative example of \textit{in situ} temperature measurements demonstrating the QSS under which the experiments are analyzed is discussed in Appendix~\ref{sec:appendix_B} and shown in Fig.~\ref{fig:Appendix_fig1}.

\begin{figure}
\centering\includegraphics[width=1\linewidth]{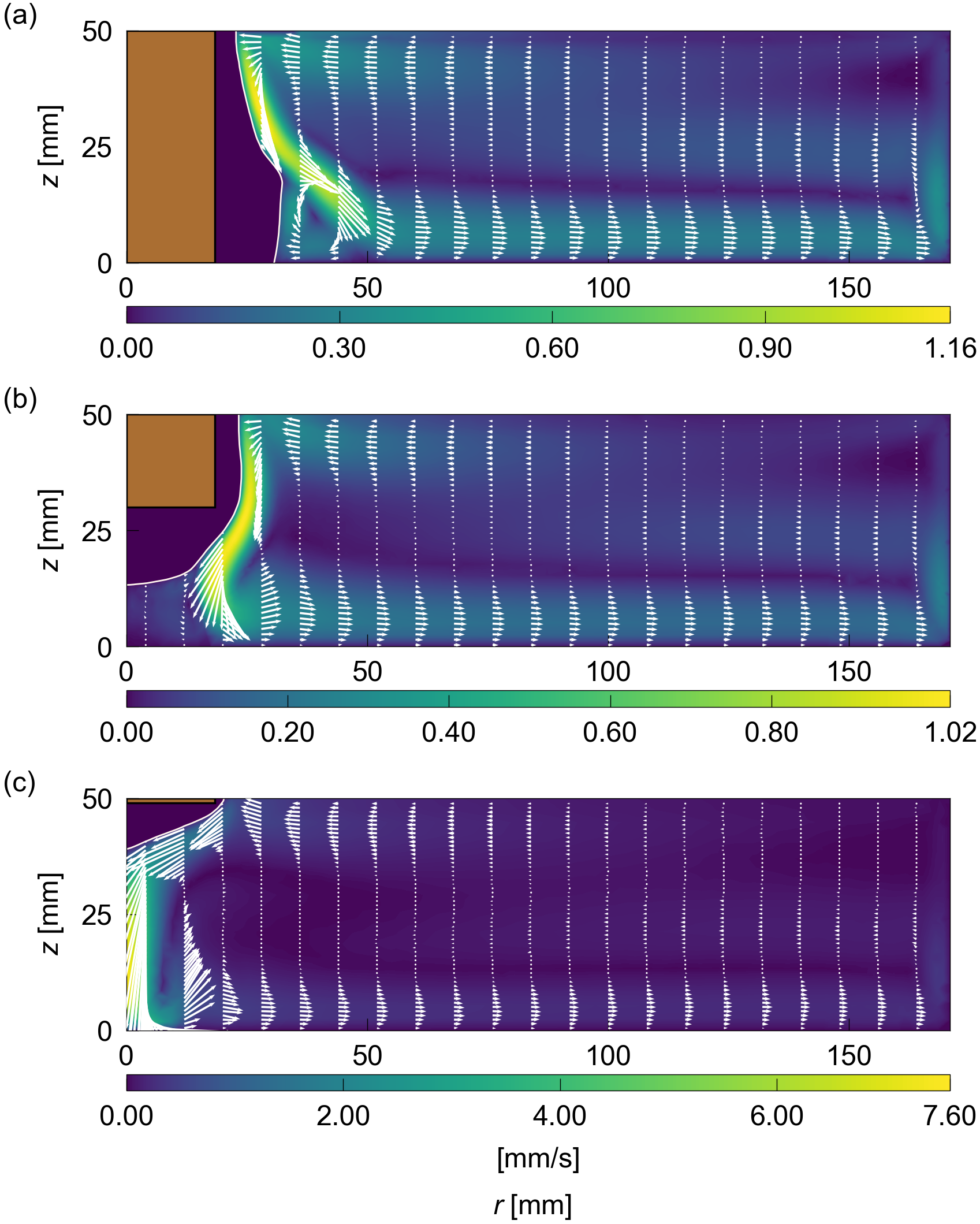}
  \caption{Interpolated velocity field for three different configurations. (a) full-depth configuration, in which ice reaches the tank bottom; (b) intermediate-depth configuration, where ice bottom reaches some intermediate depth; and (c) surface configuration, where ice remains at the air–water interface. The copper-colored region in each panel denotes the location of the temperature-controlled copper tube within the imaged domain.} 
  \label{fig:3}
\end{figure}

\subsection{Phase segmentation and interpolated velocity}

We first segment the ice and liquid phases, compute the liquid velocity field with PTV, and interpolate the vectors onto an unstructured finite-element mesh. Figure~\ref{fig:3} shows the resulting interface geometry and interpolated velocity fields for the three ice--water configurations. Together, the cases isolate how ice draft and interface shape control the organization and intensity of the convective circulation.

In the full-depth configuration (Fig.~\ref{fig:3}a), the flow organizes into two coherent cells. 

\textcolor{black}{When the ice spans the full depth, the basin behaves like a two-gear convective engine. (Fig.~\ref{fig:3}a),} A basin-scale primary convective cell fills most of the liquid domain, while a compact secondary cell remains attached to the lower ice--water interface. The fastest velocities of about $1~\textrm{mm/s}$ occur in a narrow descending jet immediately downstream of the ice front. This jet provides the main momentum pathway: after impinging on the bottom boundary, it bifurcates, with one branch feeding the secondary cell beneath the ice and the other sweeping across the lower domain, rising along the outer wall, and returning toward the ice to close the primary overturning loop. The interpolation preserves both the sharp jet gradients and the weaker recirculating flow away from the interface.

\begin{figure*}[ht]\centering\includegraphics[width=1\linewidth]{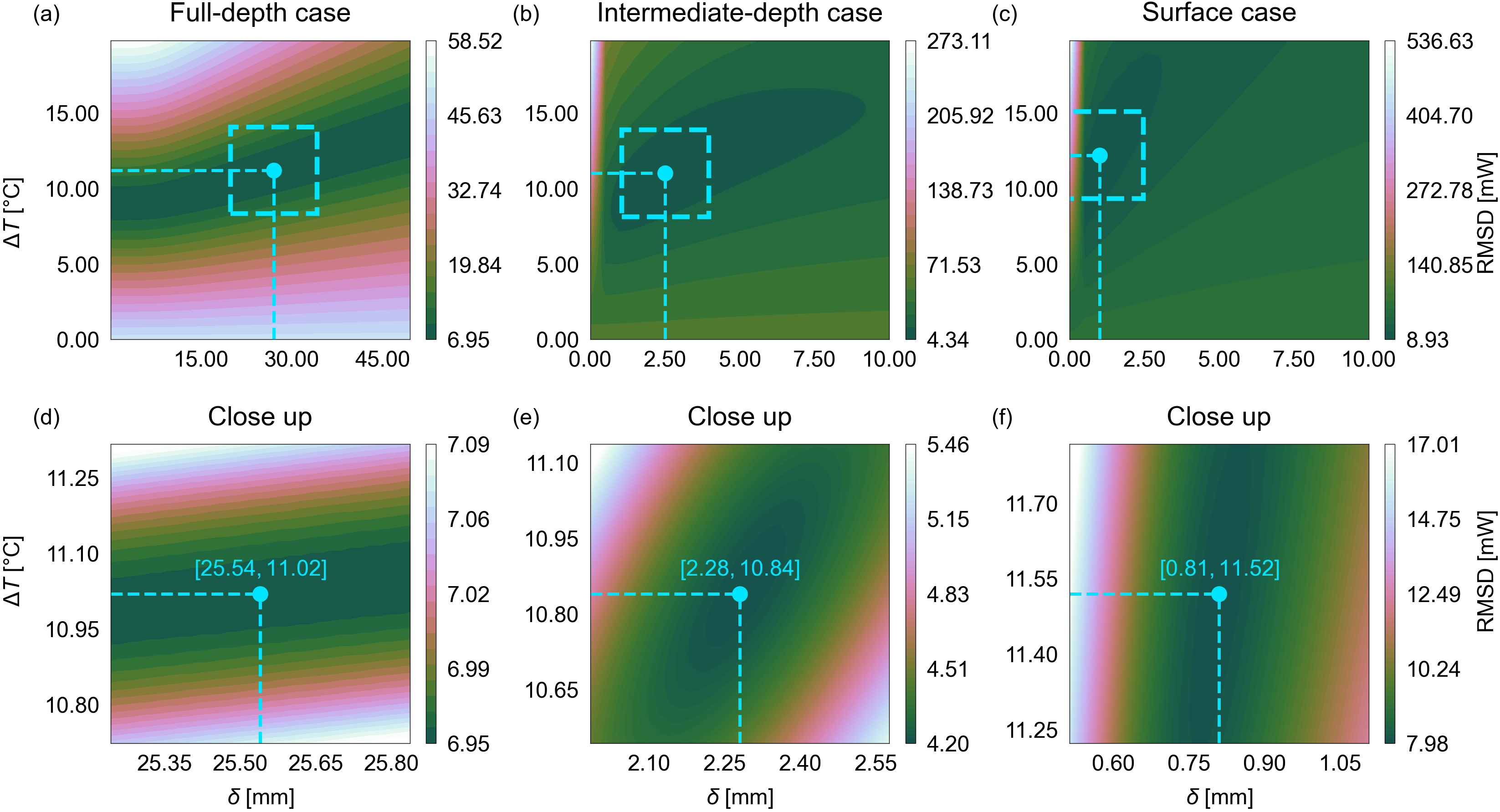}
  \caption{Two-round parameter search for the optimal parameter pair ($\Delta T$, $\delta$) based on minimization of the RMSD between local heat fluxes. Panels (a)-(c) show the first round search over the comprehensive parameter domain for the full-depth, intermediate-depth, and surface cases, respectively. Preliminary optimums occur at (a) $(\Delta T, \delta) = (27.01,\,11.20)$, (b) $(\Delta T, \delta) = (2.51,\,11.00)$, (c) $(\Delta T, \delta) = (1.01,\,12.20)$. Panels (d)-(f) show the second-round refined search over a reduced neighborhood around the preliminary optimums. Cyan markers indicate the parameter pair corresponding to the minimum RMSD. Specifically, the optimal coordinates are indicated within each subfigure.} 
  \label{fig:4}
\end{figure*}

At intermediate ice draft (Fig.~\ref{fig:3}b), the same two-cell dynamics persists, but the ice geometry shifts how the two cells share the domain. The primary circulation spans most of the water column, whereas a weaker secondary cell forms beneath the ice. The fast-moving descending jet remains partially attached to the curved ice boundary and reaches velocities of up to $1~\textrm{mm/s}$ before detaching and impinging on the bottom boundary, where it splits into the same two branches observed in the full-depth case. The segmented interface shows a smooth, monotonic curvature; its nearly vertical portion aligns with the jet trajectory, while its more horizontal portion bounds the secondary recirculation below the ice. The interpolated field captures this geometry-induced organization, including the concentrated shear within the descending jet.

When the ice is confined to the upper water column (Fig.~\ref{fig:3}c), the circulation simplifies to one dominant overturning cell that occupies nearly the entire liquid domain. 
A thicker descending jet forms downstream of the lower ice edge and initially follows the ice contour before detaching toward the bottom. After impingement, the flow turns radially outward along the base, rises near the outer wall, and returns along the upper domain toward the ice. The maximum velocity exceeds $7\ \mathrm{mm/s}$ within the descending branch. Because the ice--water interface is nearly linear and confined to the upper water column, the flow lacks the secondary recirculation found in the deeper-draft cases, but retains a coherent jet-driven overturning structure.

Overall, the phase segmentation and PTV-based interpolation preserve the dynamically relevant geometry and kinematics across all cases: ice-boundary shape, near-interface jet structure, and convective cells. These fields provide the velocity input required for the temperature reconstruction and retain the flow features that control heat transport near the ice--water interface.

\subsection{Optimal parameters and global heat budget error}

We next determine the optimal parameters $\Delta T$ and $\delta$ that define the surface-temperature distribution $T_{\rm iw}(r)$ required to close the system's heat budget, and quantify the associated global heat budget error.

The admissible ranges of $\Delta T$ and $\delta$ are constrained by \textit{in situ} temperature measurements. The temperature amplitude $\Delta T$ is bounded by the freezing temperature, $0^{\circ}\mathrm{C}$, and the measured ambient air temperature. The length scale $\delta$ is estimated from the radial distance over which the near-surface temperature approaches a plateau away from the ice--water interface. Practically, this range is obtained by fitting Eq.~\ref{eq:Taw_radial} to thermistor measurements near the air--water interface (Appendix~\ref{sec:appendix_B}; Fig.~\ref{fig:Appendix_fig2}).

Within these bounds, we perform the two-stage parameter search described in Sect.~\ref{sec:param_opt}. First, $\Delta T$ and $\delta$ are sampled on a coarse grid with resolutions of $0.2^{\circ}$C and 0.2~mm, respectively, with $\Delta T$ spanning its admissible range and $\delta$ spanning $[0,3\delta]$. This initial grid is intentionally coarser than the temperature-sensor resolution and the finest finite-element mesh spacing. For each parameter pair, we solve the advection--diffusion heat equation \eqref{eq:heat_cyl_liquid} and compute the RMSD in Eq.~\ref{eq:RMSD_heat_budget} from the local ice--water heat-flux mismatch. Second, we repeat the search in a smaller neighborhood around the preliminary optimum using finer resolutions of $0.02^{\circ}$C and 0.02~mm. This refined grid yields the final optimal values of $\Delta T$ and $\delta$.

\begin{figure*}\centering\includegraphics[width=1\linewidth]{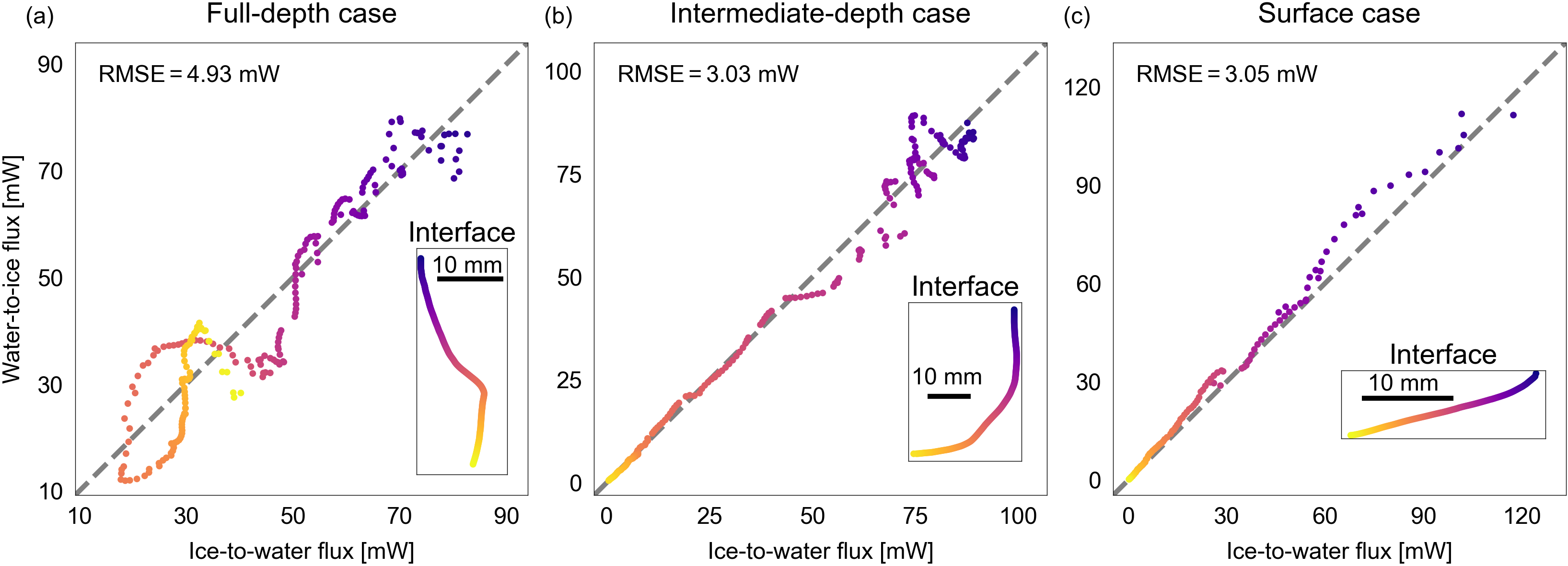}
  \caption{Comparison of segment-by-segment heat fluxes across the ice-water interfaces for the (a) full-depth, (b) intermediate-depth, and (c) surface-ice configurations. The abscissa is the ice-to-water heat flux obtained in the solid phase, and the ordinate is the water-to-ice heat flux obtained in the liquid phase through the optimization process. The color scale denotes the arc length along the ice--water interface, measured from the ice bottom, as illustrated by the inset in each panel. The dashed line denotes the identity, corresponding to perfect local heat-flux balance.} 
  \label{fig:5}
\end{figure*}

Figure~\ref{fig:4} shows the resulting RMSD maps. The top panels show the coarse search for the three configurations, and the bottom panels show the refined search around each preliminary optimum. In all cases, refinement lowers the RMSD, with final values of order $4$--$8~{\rm mW}$.

\subsection{Local error quantification}

We further assess the temperature reconstruction quality by quantifying the local heat-flux mismatch across the ice--water interface. Figure~\ref{fig:5} compares the water-to-ice and ice-to-water heat fluxes, segment by segment, along the reconstructed interface. The dashed line denotes exact local balance, and the color indicates the distance of each segment from the air--water interface. 

We quantify the local mismatch using the root mean square error (RMSE),
\begin{equation}\label{eq:RMSE}
{\rm RMSE} = \left[\frac{1}{N} \sum_{i=1}^{N} d_{\rm i}^2\right]^{1/2},
\end{equation}
where $N$ is the number of interface segments and $d_{\rm i}$ is the minimum distance from each point to the identity line \citep{pallavi2022comprehensive}.

Across the three configurations, the local heat fluxes cluster near the identity line, indicating that the reconstructed temperature fields largely satisfy interfacial thermal balance. The full-depth and intermediate-depth cases show good pointwise agreement, with RMSE values of $4.93$~mW and $3.03$~mW, respectively (Fig.~\ref{fig:5}a,b). In both cases, the heat-flux magnitude generally increases upward along the ice--water interface. The surface-ice case spans a wider range of heat fluxes and shows lower local deviations near the upper interface, yielding an RMSE of $3.05$~mW (Fig.~\ref{fig:5}c).

We also compare the residual interface displacement implied by the optimized heat-budget error with the displacement measured from image segmentation. Using the minimum RMSD from Fig.~\ref{fig:4} over a conservative QSS window of $\tau=6$~h, we estimate $\ell_{\rm res}$ from Eq.~\ref{eq:l_res} and compare it with the image-based variability $\ell_{\rm exp}$ from Eq.~\ref{eq:l_exp}. For the full-depth and intermediate-depth cases, $\ell_{\rm res}=0.06~{\rm mm}$, which is smaller than $\ell_{\rm exp}=0.2~{\rm mm}$. Thus, the residual heat imbalance would produce an interface displacement below the detection uncertainty. For the surface-ice case, $\ell_{\rm res}=0.4~{\rm mm}$ exceeds $\ell_{\rm exp}=0.08~{\rm mm}$, although both remain submillimetric. This larger residual is physically consistent with the surface-confined geometry. In this case, a small ice--water interface must accommodate heat input from a comparatively large air--water surface, producing the largest interfacial heat-transfer rate among the three configurations. In addition, the ice is located near the free surface, where laboratory thermal fluctuations are the largest. These factors amplify the sensitivity of the heat budget to boundary-condition uncertainty and explain the larger mismatch.

Overall, the local and global error checks show that the reconstructed temperature fields preserve interfacial heat-flux balance to within a few milliwatts. This accuracy is sufficient to resolve the spatial structure of near-interface heat transport and provides a reliable basis for estimating net water-to-ice heat flux.

\subsection{Results: two-phase temperature reconstruction}

\begin{figure}
\centering\includegraphics[width=1\linewidth]{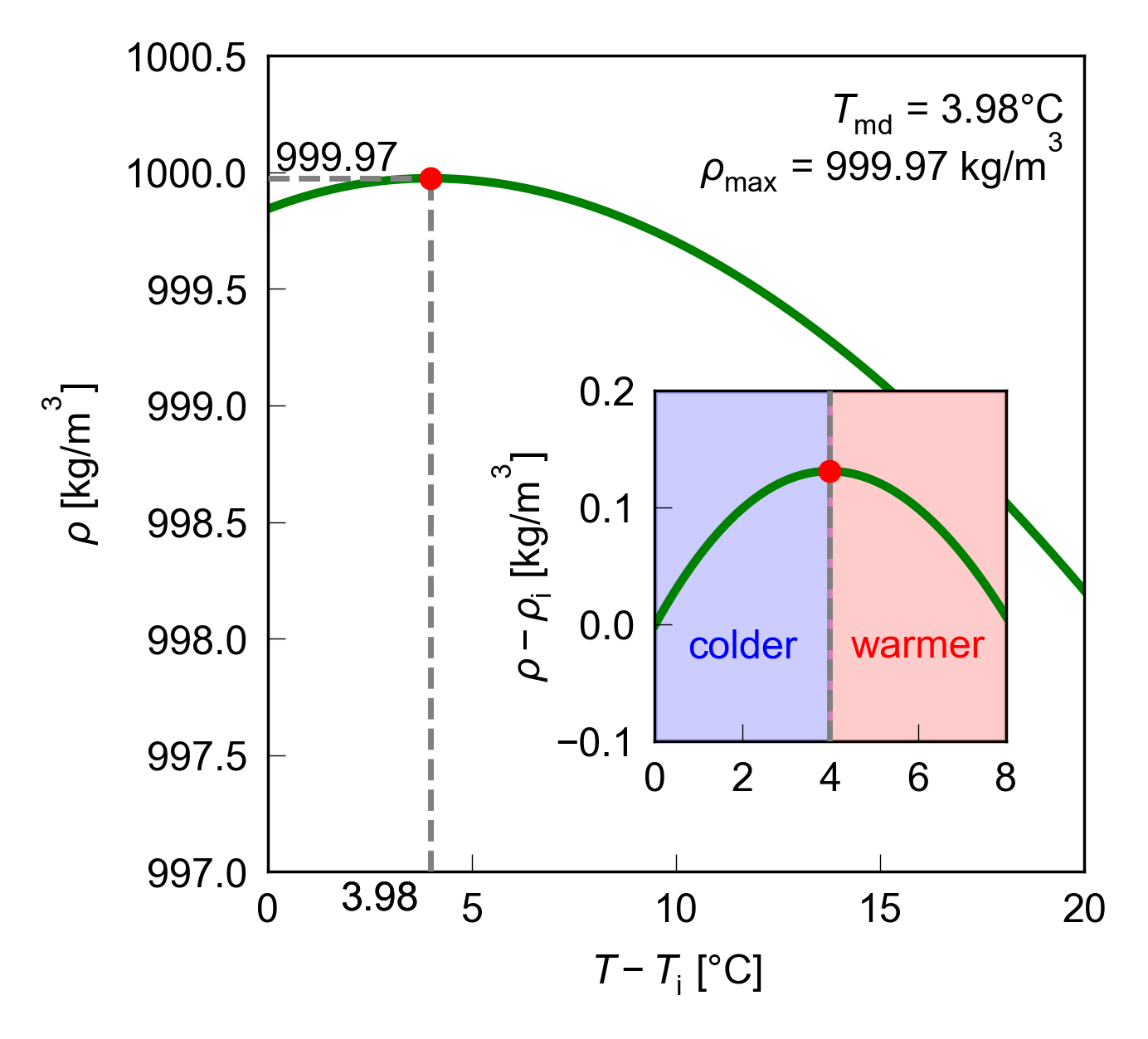}
  \caption{Equation of state for freshwater showing the non-monotonic dependence of density on temperature. $\rho_{\rm i}$ denotes the water density at the freezing point. The density increases monotonically with temperature from $0^{\circ}\mathrm{C}$ to the temperature of maximum density, $T_{\rm md}\approx3.98^{\circ}\mathrm{C}$, where density reaches $\rho_{max}\approx999.97\mathrm{kg\,m^{-3}}$. Above $T_{\rm md}$, density decreases with further increasing temperature.}
  \label{fig:6}
\end{figure}

\begin{figure}
\centering\includegraphics[width=1\linewidth]{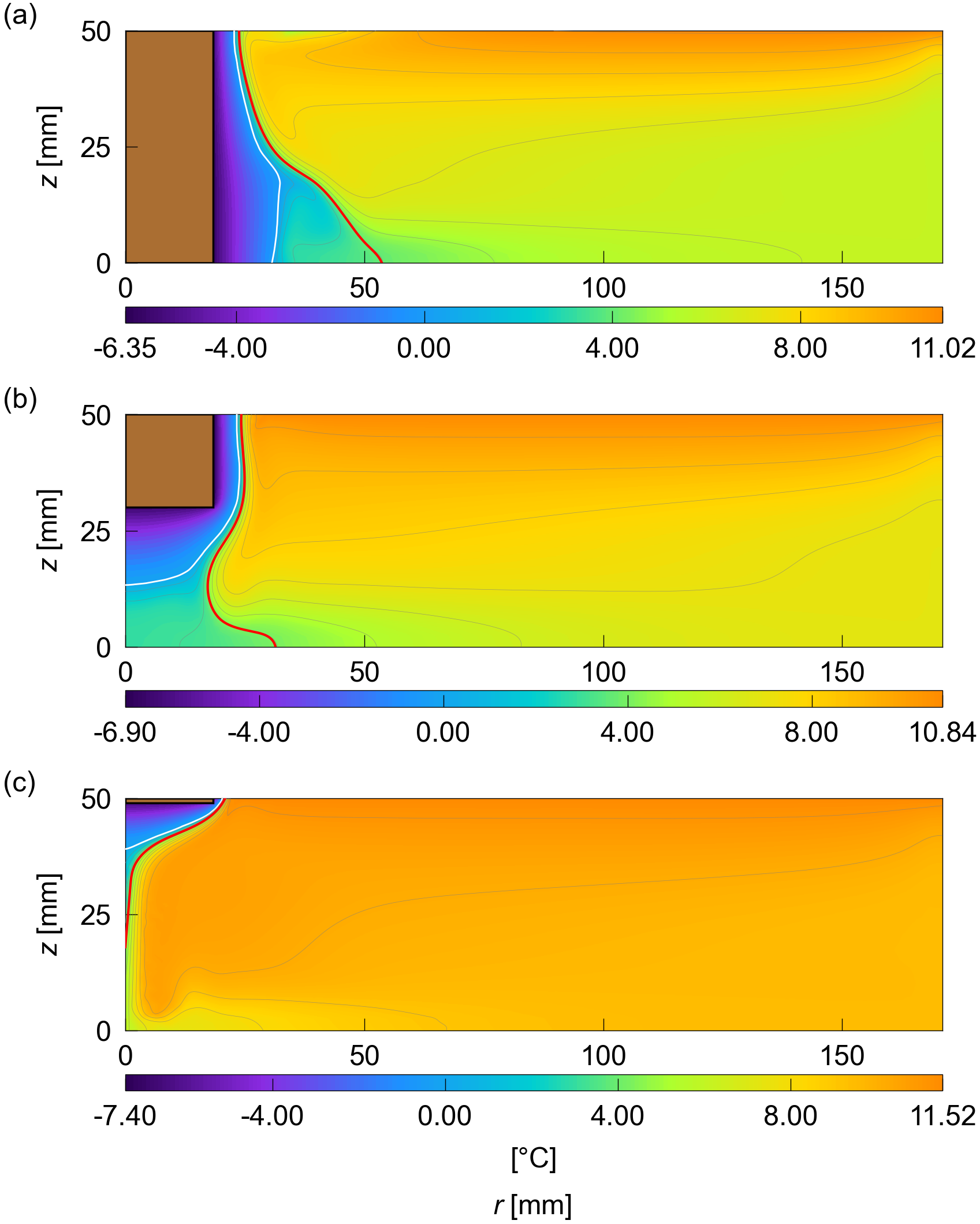}
  \caption{Two-phase temperature reconstruction in a radial--vertical plane for the three ice configurations: (a) full-depth ice, for which the ice reaches the tank bottom; (b) intermediate-depth ice, for which the ice extends to an intermediate depth; and 
(c) surface ice, for which the ice remains near the air--water interface. 
The brown region indicates the copper core in the axisymmetric system. 
The black and white contours denote the copper--ice and ice--water interfaces, respectively, with the latter at the freezing temperature. 
The red contour marks the temperature-of-maximum-density isotherm, $T_{\rm md}=3.98^{\circ}\mathrm{C}$, and gray contours indicate isotherms between $1^{\circ}\mathrm{C}$ and $10^{\circ}\mathrm{C}$.} 
  \label{fig:7}
\end{figure}

The validated two-phase reconstructions span temperatures from approximately $-8^{\circ}\mathrm{C}$ to the freezing point, $T_{\rm i}=0^{\circ}\mathrm{C}$, within the ice and from $T_{\rm i}$ to approximately $12^{\circ}\mathrm{C}$ within the liquid (Fig.~\ref{fig:7}). The liquid therefore encompasses the anomalous thermodynamic regime of freshwater associated with its non-monotonic equation of state (EOS) (Fig.~\ref{fig:6}). Between $T_{\rm i}$ and the temperature of maximum density, $T_{\rm md}\approx3.98^{\circ}\mathrm{C}$, water density increases with temperature, whereas for $T>T_{\rm md}$, density decreases with temperature. Consequently, mixing water parcels whose temperatures lie on opposite sides of $T_{\rm md}$ can produce a mixture that is denser than that predicted by linear averaging and, for suitable parent temperatures and mixing proportions, even denser than either parent parcel. The resulting negatively buoyant mixture can sink, thereby reinforcing convective motion Fig.~\ref{fig:6}. This nonlinear mixing process, known as cabbeling, contributes to the coupled temperature and velocity fields shown in Fig.~\ref{fig:3}.

Across the three configurations, the temperature reconstructions reveal a systematic correspondence with the measured flow structure. In the liquid phase, the $T_{\rm md}$ isotherm, shown by the red contour in Fig.~\ref{fig:7}, closely follows the descending branch of the circulation adjacent to the ice--water interface. This branch coincides with the cabbeling-driven downwelling identified in the velocity field (Fig.~\ref{fig:3}). The downwelling remains attached to the ice boundary over much of its path and separates near the lower part of the ice--water interface. This separation partitions the liquid into two thermodynamic regions: a colder region with $T_{\rm i}<T<T_{\rm md}$ and a warmer region with $T>T_{\rm md}$.

The full-depth and intermediate-depth ice configurations exhibit a two-cell circulation structure (Figs.~\ref{fig:3}a,b and \ref{fig:7}a,b). In both cases, the descending cabbeling-driven current reaches the lower part of the liquid domain and bifurcates. One branch turns toward the ice and feeds a compact recirculation cell beneath the ice, trapping the coldest liquid water, with temperatures between $T_{\rm i}$ and $T_{\rm md}$. The other branch moves away from the ice along the bottom boundary and drives a larger overturning circulation across the radial--vertical plane. This larger cell resembles a horizontal-convection-like circulation: relatively warm water near the outer air--water interface descends toward the ice contact region, is cooled near the ice, and then participates in the cabbeling-driven returning flow.

By contrast, the shallow surface-ice configuration produces a simpler circulation pattern (Figs.~\ref{fig:3}c and \ref{fig:7}c). Here, the cabbeling-driven downwelling develops closer to the axis of the cylindrical domain and spreads radially along the base, generating one dominant overturning cell rather than a two-cell structure. The warmest liquid water remains near the free surface, and the reconstructed field shows a weaker vertical temperature stratification than in the deeper-ice cases.

The solid-phase temperature field is controlled by conduction through the ice and by the imposed thermal conditions at the ice boundaries. In the surface-ice case, the upper ice boundary is fixed by the copper plate at the air--water interface, whereas the lower ice boundary is constrained by contact with water near the freezing point. The reconstructed temperature, therefore, varies monotonically across the ice thickness, consistent with conductive heat transfer between the copper-controlled boundary and the ice--water interface. 

Overall, these results show that the inverse reconstruction captures both the two-phase thermal structure and its dynamical coupling to the flow. In particular, the reconstructed temperature fields link freshwater cabbeling, circulation pattern, and interfacial heat exchange within a single experimental framework. They also agree with simulations in which cabbeling generates a descending plume along the ice--water interface, drives two-cell circulation, and produces comparable interface morphologies \citep{bourdillon2015numerical,yang2022abrupt}. Related studies have reported similar freshwater dynamics associated with the non-monotonic EOS, but without resolving the strong two-phase thermal coupling demonstrated here \citep{wei1994density,ishikawa2000numerical,noto_PRF_2026}. By linking velocity, temperature, and ice geometry, this framework provides a direct experimental route to quantify heat transport and to test simulation results.

\subsection{Reconstruction vs \textit{in situ} measurements}

Finally, we validate the optimized liquid-phase temperature reconstruction against \textit{in situ} thermistor measurements acquired at different depths and radial positions. Because of the axisymmetric flow structure, the temperature measurements were placed at the same $(r,z)$ coordinates used for reconstruction but at different azimuthal locations from the PTV plane (Fig.~\ref{fig:1}). This arrangement provides an independent pointwise test of the reconstructed thermal field without constraining the inverse problem.

\begin{figure*}\centering\includegraphics[width=1\linewidth]{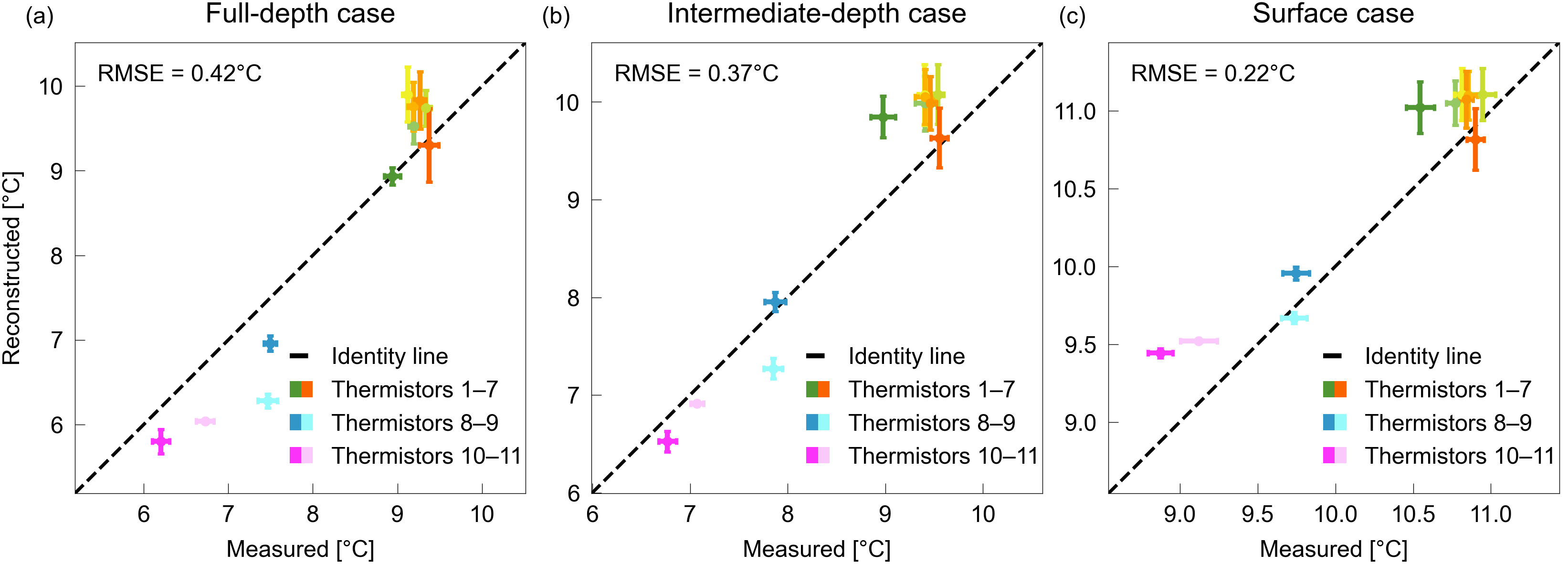}
  \caption{Comparison between measured and reconstructed temperatures for the (a) full-depth, (b) intermediate-depth, and (c) surface-ice configurations. The dashed line denotes the identity, corresponding to perfect agreement between measurements and reconstruction. Thermistor locations are shown in Table ~\ref{table:Table_1}. Thermistors 1--7 are positioned near the air--water interface, at the same depth and different radial locations. Thermistors 8--9 are positioned at the same intermediate depth and two radial locations, whereas thermistors 10--11 are positioned at the same vertical distance to the bottom boundary and two radial locations.
} 
  \label{fig:8}
\end{figure*}

Owing to the thermistors' size, only a limited number of measurement locations were accommodated within the water domain. 
Temperature was measured at 11 locations (Fig.~\ref{fig:1}, Table~\ref{table:Table_1}). Seven thermistors were positioned $5~\mathrm{mm}$ below the air--water interface and distributed radially. Two thermistors were placed at intermediate depth, $25~\mathrm{mm}$ below the air--water interface, with one near the domain center and one near the outer annular boundary. The remaining two thermistors were located $5~\mathrm{mm}$ above the tank bottom, again at interior and outer radial positions. This layout samples the near-surface thermal boundary layer, the interior water column, and the near-bottom region.

Figure~\ref{fig:8} compares the time-averaged thermistor measurements with the reconstructed temperatures at each sensor location for the three configurations. Each reconstructed value is locally averaged over a circular region with diameter three times the sensor-tip diameter, $2~\mathrm{mm}$, to account for finite sensor size and positioning uncertainty. Error bars denote temporal variability in the measurements and spatial variability in the reconstructed field over the averaging region. The dashed line indicates the identity. Agreement is quantified using the RMSE, computed across the $N=11$ sensors in each configuration using Eq.~\ref{eq:RMSE}.

Across the three configurations, the reconstructed temperatures generally follow the identity line, indicating good pointwise agreement with the independent measurements. Near-surface sensors (locations 1--7), located close to the air--water interface, show the closest agreement but the largest variability, consistent with strong thermal gradients and sensitivity to the inferred surface boundary condition. Deeper sensors (locations 8--11) show smaller variability but larger systematic deviations. In the full-depth and intermediate-depth cases, reconstructed temperatures are slightly colder than the measurements at depth. In the surface-ice case, the reconstruction agrees well near the surface but slightly overestimates deeper temperatures.

Overall, the validation indicates that the reconstruction captures the measured thermal structure across all three ice--water configurations. The remaining discrepancies are primarily systematic rather than stochastic, especially in the deeper water column. The RMSE ranges from $0.22^{\circ}\mathrm{C}$ in the surface-ice case to $0.42^{\circ}\mathrm{C}$ in the full-depth case, remaining small relative to the experimental temperature range. The reconstructed fields are therefore sufficiently accurate for estimating interfacial heat fluxes along complex ice--water boundaries.


\section{Discussion}\label{sec:discussion}

\subsection{Remarks on method's requisites}

The method relies on five experimental requisites. First, the system involves a two-phase freshwater configuration in which liquid water and ice are in direct contact, and both phases have non-isothermal temperature fields. The ice-water and copper-ice interfaces must be identified because they define the geometry on which the thermal boundary conditions and interfacial heat fluxes are imposed and evaluated.

Second, the leading-order dynamics must be representable in a quasi-two-dimensional plane. In the present annular system, this requires weak azimuthal variability, so that a single radial--vertical plane captures the dominant advective and conductive heat-transfer pathways. This condition is essential because the inverse problem is solved in the same plane where velocity and phase boundaries are measured.

Third, the experiment must reach a quasi-steady state before reconstruction. For this, we must minimize: temperature fluctuations in the atmosphere, heat exchange at the lateral and bottom boundaries, and laser exposure during velocity measurements. Statistical stationarity allows the temperature field to be constrained by heat-budget closure and enables the unknown air--water temperature distribution to be inferred from energy balance rather than imposed \textit{a priori}.

Fourth, the liquid velocity field must be independently measured and sufficiently resolved, with approximately $O(10^{2})$ velocity measurements per millimeter square, and interpolated onto a computational mesh of around $O(10^{4})$ nodes. These resolution requirements are not universal but depend on the characteristic spatial and temporal scales of the flow under investigation. The choice of the imaging system and acquisition parameters should therefore be guided by the characteristic properties of the flow.
The reconstruction treats this velocity field as a prescribed advective input, not as an unknown. The measurements must therefore resolve the principal flow structures, including descending cabbeling-driven jets, recirculation zones and overturning cells. Poor velocity resolution would directly affect the advective heat transport term and propagate into errors in the reconstructed temperature field and heat-flux estimates.

Fifth, all boundary conditions must be known or physically constrained; otherwise fully predefined. The copper-ice boundaries are fixed by the imposed copper-core temperature, and the ice-water interface is constrained by the freezing temperature. If the air-water boundary condition is not prescribed, it can be recovered through an inverse optimization that minimizes the residual heat-budget imbalance across the coupled ice-water system. 

Independent thermistor measurements are acquired in a different radial--vertical plane (Fig.~\ref{fig:1}) and used only after the reconstruction is complete. They therefore provide an independent validation of the reconstructed temperature field without over-constraining the inverse problem.

By design, the thermistors are small enough to avoid leading-order hydrodynamic disturbance. Using a thermistor diameter $d_p\simeq 2~{\rm mm}$ and a characteristic horizontal velocity $U=O(1)~{\rm mm\,s^{-1}}$, the probe Reynolds number is
$$Re_p=\frac{U d_p}{\nu}\simeq O(1),$$
for near-freezing freshwater with $\nu\simeq 1$--$2\times10^{-6}~{\rm m^2\,s^{-1}}$. This lies well below the onset of periodic vortex shedding behind a circular cylinder, $Re_p\simeq 47$--$50$, and below the three-dimensional wake-transition regime, $Re_p=O(10^2)$ \citep{roshko1954,Williamson1996vortex}. Thus, the thermistors may generate only weak, steady viscous disturbances localized near the probes, not wake structures capable of altering the large-scale convective circulation.

\subsection{Remarks on method limitations and opportunities for future developments}

The requisites discussed above make the two-phase reconstruction physically tractable; they also define its current range of validity. The main limitations concern temporal evolution, validation strategy, material-property assumptions, and background thermal bias.


First, the system is not perfectly steady. Despite enclosing the setup in a 50~mm-thick XPS box, with an additional 50~mm layer around the tank sides and bottom, small diurnal temperature fluctuations persist in both air and water. At QSS, water-temperature fluctuations are approximately $\pm 0.05^{\circ}$C (Fig.~\ref{fig:Appendix_fig1}). Velocity measurements can introduce additional heat exchange because optical access requires a narrow opening ($\approx 1$~cm) in the tank wall facing the laser and a fully open wall facing the camera. Thus, wall-opening time and laser exposure must be minimized. Automation reduces, but does not fully eliminate, this spurious thermal forcing during velocity acquisition.

Second, as we investigate QSS regimes, our temperature reconstruction is independent of the initial condition. This choice is appropriate here because the goal is to relate ice geometry, circulation structure, and heat transfer after the system reaches statistical stationarity. However, many natural and laboratory ice-water systems feature evolving phase geometry, unsteady forcing and transient convective dynamics \citep[e.g.,][]{wang2021equilibrium,wang2021growth,mccutchan2023experimental,perry2025rotation,johnson2025shape,noto2026melting}. Experiments with controlled initial conditions could extend the framework toward time-dependent reconstruction of temperature \citep{nakao2025reconstruction}, phase boundary and liquid motion.

Third, reconstruction quality is assessed through heat-budget closure and independent \textit{in situ} temperature measurements. The first provides an integral consistency check, whereas the second validates the local liquid temperature field. Because thermistors must be inserted into the liquid, they may perturb the flow if placed near sensitive regions, especially cabbeling-driven descending jets. Although the observed circulation remains broadly non-turbulent, probes should be kept away from dominant advective pathways. Future implementations could reduce this limitation through smaller sensors, optimized probe geometry, or non-invasive temperature diagnostics \citep[e.g.,][]{smith2001temperature,anders2020simultaneous}.

Fourth, the reconstruction assumes constant material properties within each phase. The measured velocity field already contains the effects of temperature-dependent density and viscosity. In contrast, thermal conductivity and heat capacity enter the heat equations as constants (Eqs.~\ref{eq:heat_cyl_liquid} and \ref{eq:heat_cyl_solid}). This approximation is reasonable for the present conditions, particularly because recent cold-water convection studies show that variations in thermal conductivity have weaker dynamical effects than variations in density or viscosity \citep{noto_PRF_2026,estay2026effects}. Nevertheless, allowing thermal conductivity and heat capacity to vary with temperature would improve thermodynamic consistency and refine estimates of temperature gradients and interfacial heat flux.

Looking forward, a more demanding extension is salinity. Salinity is excluded here, but it is central to ice--ocean and glacier--fjord systems because it modifies density, buoyancy, stratification, and interfacial melting. Extending the method to saline stratified flows would require separating the contributions of temperature and salinity to buoyancy, and ideally reconstructing both scalar fields. This inverse problem is substantially harder, but it would broaden the framework from freshwater convection to thermohaline ice--water dynamics relevant to marine environments \citep[e.g.][]{fitzmaurice2017nonlinear,wilson2023double,bellincioni2025melting}.

In summary, these limitations emphasize where the present method applies and identify the most useful extensions: transient reconstruction, less intrusive validation, temperature-dependent material properties, and thermohaline coupling.

\section{Summary and outlook}\label{sec:summary}

We reconstruct the two-phase temperature of ice-water systems open to the atmosphere using a single framework. The framework combines kinematic measurements, sparse thermal constraints, and a physics-constrained inverse formulation for heat transport. The method extends previous velocity-based approaches for single-phase buoyancy-driven flows to coupled solid--liquid systems \citep{noto2023reconstructing}. It does so by solving advection--diffusion in the liquid water and conduction in the ice in separate subdomains, subject to prescribed as well as \textit{a priori} unknown air--water interface boundary conditions. This enables the reconstruction of temperature fields on both sides of irregular, non-isothermal ice's boundaries, including the interfacial thermal gradients that control heat exchange, ice growth, and melting. By operating under laboratory conditions with free surfaces and complex ice geometry, while avoiding dyes, refractive-index matching, and dense intrusive thermometry, the approach provides a practical and minimally invasive methodology for quantifying thermo-fluid coupling in evolving ice--water environments.

The methodology developed here can be extended to a broad range of ice--water systems. In Earth science, potential applications include water interacting with floating ice bodies, such as icebergs and ice floe; laterally ice-bounded environments, such as proglacial lakes and supraglacial meltwaters; and partially or fully ice-covered water bodies \citep[e.g.,][]{hester2021aspect,weady2022anomalous,mccutchan2023experimental,johnson2025shape,perry2025rotation,wolterman2025wave,bellincioni2025melting,noto2026melting,kirillin2015axisymmetric,bouffard2019,2026_estay_pnas_nexus,wells2008circulation,rabaux2026hysteretic,mcgrath2026rapid}. Although these systems differ in geometry, forcing, and scale, they share coupled heat transport, buoyancy-driven circulation, and phase change at an evolving ice--water interface. More broadly, the framework may also support studies of food processing, thermal-energy storage, and industrial cooling systems in which coexisting solid and liquid phases regulate thermal performance \citep[e.g.,][]{goldstein1979heat,stickland2007experimental,kauffeld2010ice,lilley2021phase}.

In such two-phase water systems, temperature is not merely a diagnostic quantity; it is the state variable that couples buoyancy generation, liquid circulation, and solid-phase growth or retreat. By reconstructing temperature simultaneously in water and ice, the method enables quantification of conductive heat transport within the solid, advective--diffusive heat transport within the liquid, and heat exchange across the ice--water interface. When combined with measurements of interface geometry and liquid velocity, these fields can be used to test scaling relationships among ice geometry, circulation structure, and interfacial heat flux. The method therefore provides controlled laboratory constraints for processes that are otherwise investigated primarily through field observations, numerical simulations, or more idealized experiments. Such constraints are increasingly important as ice-covered and ice-contact aquatic environments respond to climatic warming, while the local mechanisms governing ice loss and cold-water convection remain difficult to observe directly \citep{du2024physics}. Laboratory studies can therefore isolate the thermofluid mechanisms governing ice--water evolution and help guide field-scale investigations in natural environments where direct measurements remain limited.

\appendix

\section{Control experiments}\label{sec:appendix_A}

We performed control experiments to assess repeatability, quantify background thermal drift, and test whether laser-induced heating could explain the observed flow and reconstructed temperature fields. The experimental protocol, imaging sequence, thermistor acquisition, and environmental enclosure were kept unchanged, but the ice-imposed thermal boundary condition was removed. \textcolor{black}{In this no-ice configuration, the temperature remained nearly uniform at approximately $17.3^{\circ}$C, with differences among sensors limited to $\pm 0.1^{\circ}$C.} \textcolor{black}{This thermal state differs substantially from the ice-forced cases, where temperatures ranged from $0^{\circ}$C to approximately $10^{\circ}$C (Fig.~\ref{fig:7}).} Measured motions in the no-ice case were weak, with a median velocity magnitude of 0.04~$\rm mm,s^{-1}$, about one order of magnitude smaller than the velocities observed in the ice-forced cases (Fig.~\ref{fig:3}). These residual motions were likely driven by one-sided laser absorption and imperfectly adiabatic boundaries. \textcolor{black}{Although this weak background motion is not negligible, its spatial structure and thermodynamic state differ strongly from those of the ice-forced cases and cannot account for the coherent thermal structures or circulation patterns observed under ice forcing.} In particular, cabbeling-driven dynamics appeared only when the ice--water thermal boundary condition was imposed. These controls therefore bound the contribution of ambient drift and confirm that the dominant response is driven primarily by interfacial heat exchange at the ice--water and air--water boundaries, consistent with recent numerical studies \citep{bourdillon2015numerical,noto_PRF_2026}.

\section{Quasi-steady state and surface temperature}\label{sec:appendix_B}

\begin{figure*}
\centering\includegraphics[width=1\linewidth]{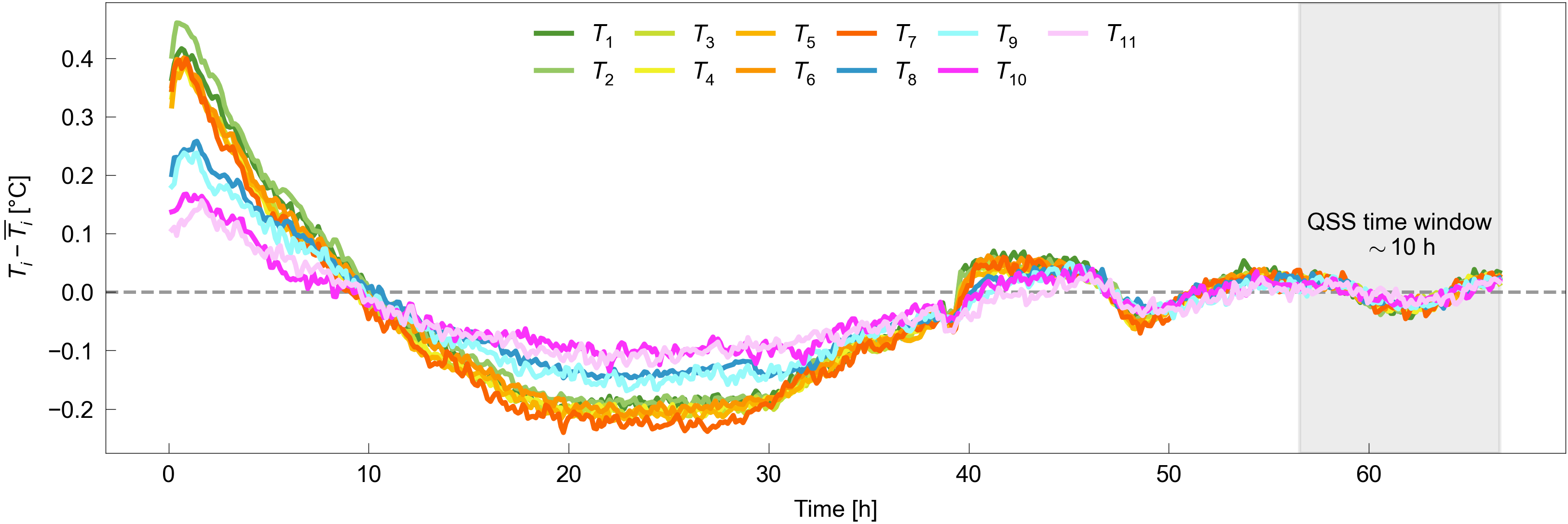}
  \caption{Temporal evolution of temperature fluctuations measured by the 11 thermistors with positions referenced from Table~\ref{table:Table_1} throughout the experiment. The shaded region denotes the QSS window. The temperature fluctuations converge toward zero across all measurement locations.}
  \label{fig:Appendix_fig1}
\end{figure*}

We assessed QSS conditions using \textit{in situ} temperature measurements and ice--water interface displacement. QSS was reached after 3--5 days, depending on the experiment. Figure~\ref{fig:Appendix_fig1} shows an example of the temperature evolution recorded by the 11 sensors deployed in the liquid phase. For each $j$ sensor, we plot $T_j-\overline{T}_j$, where $\overline{T}_j$ is the mean temperature over the final 10~h. At early times, $T_j-\overline{T}_j$ exhibits a transient phase with large sensor-to-sensor variability, reflecting the different sensor distances from the air--water interface (Table~\ref{table:Table_1}). During the final day, however, all curves collapse, with diurnal temperature variations smaller than $\pm 0.1^{\circ}$C, demonstrating that the heat content in the system has stabilized and reached a QSS. These residual oscillations are attributed to ambient temperature fluctuations in the laboratory.

\begin{table}[h!]
\centering
\caption{Thermistor positions for all experimental trials}
\begin{tabular}{ccc}
\hline
Thermistor & $r$ position [mm] & $z$ position [mm] \\
\hline
1  & 39.5  & 45 \\
2  & 59.5  & 45 \\
3  & 79.5  & 45 \\
4  & 99.5  & 45 \\
5  & 119.5 & 45 \\
6  & 139.5 & 45 \\
7  & 159.5 & 45 \\
8  & 79.5  & 25 \\
9  & 139.5 & 25 \\
10 & 99.5  & 5 \\
11 & 159.5 & 5 \\
\hline
\end{tabular}
\label{table:Table_1}
\end{table}

The surface temperature sensors, located 5 mm below the air--water interface, together with the freezing temperature at the water--ice--air contact line, provide the constraints used to define the radial air--water interfacial boundary condition, $T_{\rm aw}$, in Eq.~\ref{eq:Taw_radial}. Figure~\ref{fig:Appendix_fig1} shows an example of the near-surface radial temperature distribution at QSS for case 1, discussed in Sect.~\ref{sec:results}. The sensor measurements are shown as markers, while the optimized surface-temperature model, obtained from the fitted hyperparameters $\Delta T$ and $\gamma$, is shown by the grey line. Because the thermistors are located 5 mm below the interface, these measurements serve as constraints on the optimization rather than direct observations of the air--water interfacial temperature.
As expected, $T_{\rm aw}(r)$ is consistently warmer than the measurements taken 5 mm below the air-water interface, except near sensor 1, where the discrepancy falls within the systematic uncertainty quantified for the experimental system, reported in Fig.~\ref{fig:8}.\\

\begin{figure}
\centering\includegraphics[width=1\linewidth]{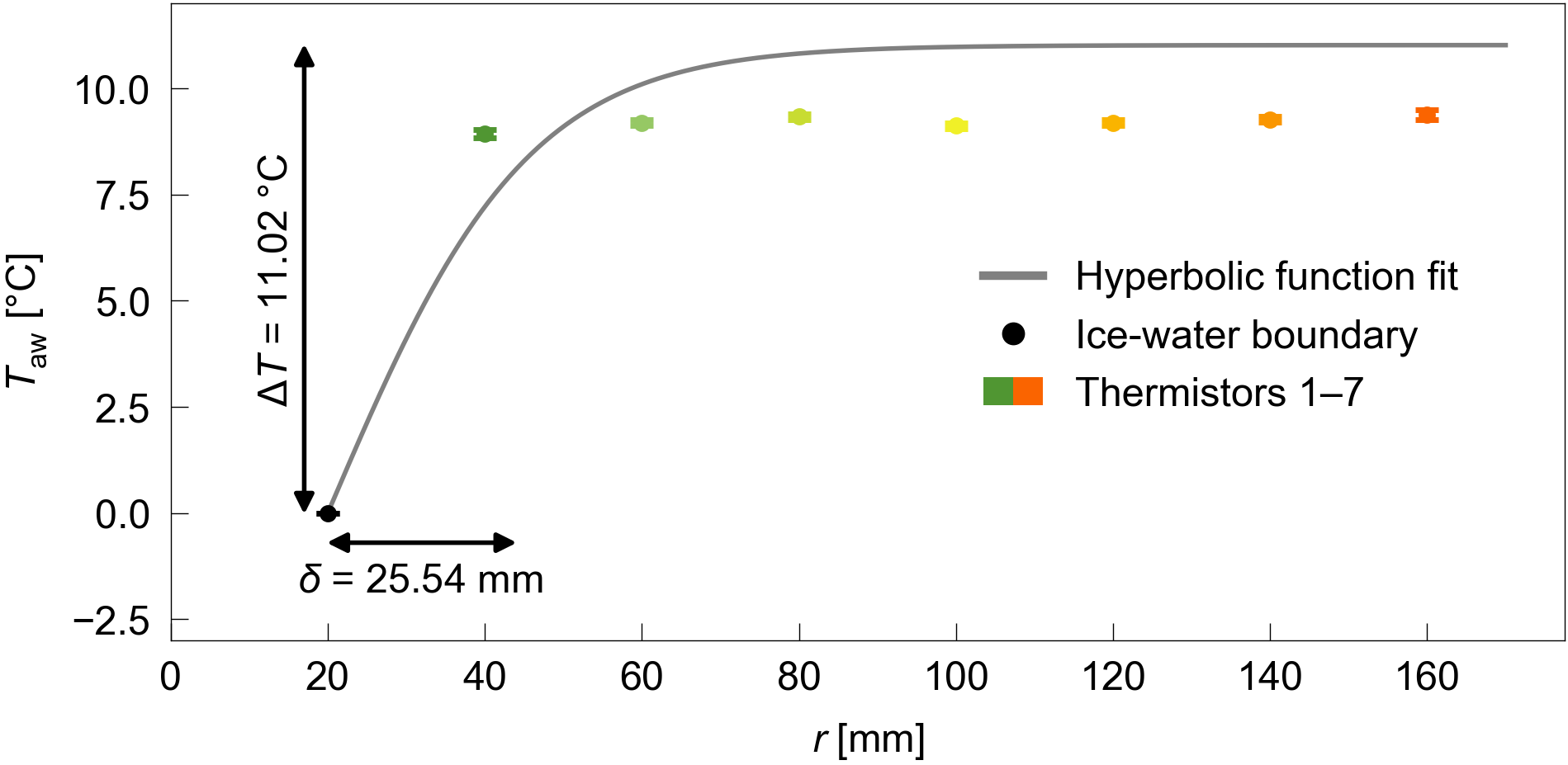}
  \caption{Radial temperature distribution for thermistors 1-7 and the corresponding hyperbolic function fit from Eq.~\ref{eq:Taw_radial}. The fitted curve represents the estimated air-water interface temperature, while the  thermistor measurements are slightly lower due to the sensor locations 5~mm below the interface ($z=45$~mm)} 
  \label{fig:Appendix_fig2}
\end{figure}

\noindent{\bf Acknowledgement} ZW acknowledges support from the University of Pennsylvania Benjamin Franklin Fellowship.\\

\noindent{\bf Author contributions} ZW contributed to conceptualization, experimental design and implementation, data acquisition and analysis, and drafting of the manuscript. DN contributed to conceptualization, experimental design, original development of the implemented PTV algorithm, mentoring, drafting of the manuscript, and funding acquisition. DJJ contributed to experimental procedure, mentoring, and manuscript editing. HNU contributed to conceptualization, mentoring, drafting of the manuscript, and funding acquisition.\\

\noindent{\bf Funding} HNU acknowledges support from the start-up grant provided by the University of Pennsylvania, and DN acknowledges support from JST EXPERT-J, Japan Grant Number JPMJEX2511.\\

\noindent{\bf Data availability} The information required to reproduce the experimental setup is provided in Sect.~\ref{sec:methods}. The finite-element solver used to integrate the heat equations in the liquid and solid phases was implemented with the open-access \href{https://github.com/FreeFem}{FreeFEM} platform.\\

\noindent{\bf Conflict of interest} The authors declare no conflict of interest.


\bibliography{references}

\end{document}